\documentstyle[12pt,twoside,epsfig,CMP209]{article}
\title[Materials of the round table ``Phase transitions and critical
phenomena: past, present, and future"] {Materials of the round
table ``Phase transitions and critical phenomena: past, present,
and future"}
\author[]{}
\address{}
\begin{document}

\maketitle

\begin{abstract}
The paper contains materials of the discussion concerning phase
transitions and critical phenomena which took place during the
Workshop on modern problems of soft matter theory (Aug. 27 -- Aug.
31, 2000, Lviv, Ukraine).
\end{abstract}
\section*{Is that true that all principal work in the phase
transitions theory has already been done up to the middle
80-ies and there is neither new physics there to be explained nor
a need to apply efforts to theoretical studies?}

\noindent{\bf Yurij Holovatch}. Dear colleagues, I have the
pleasure to open our round table during which we will have a
possibility to discuss past, present and possible future of the
field most of us are currently interested in: the phase
transitions theory. Perhaps it is hard to name another branch of
modern physics which was so essentially changed during last 30
years as the theory of phase transitions and critical phenomena.
One can say even more: having appeared in a form of different
variants of a mean field theory for describing critical points in
liquids and equilibrium phase transitions in problems of
statistical physics and thermodynamics, the theory of phase
transitions and critical phenomena now appears to be an
interdisciplinary science as, say, the theory of vibrations. Also,
the sphere of notions contained in the term "critical phenomena"
has been essentially broadened. Aside from traditional critical
points in liquids and equilibrium thermodynamic 2nd order phase
transitions, this term includes non-equilibrium dissipative phase
transitions, percolation phenomena as an example of a geometrical
phase transition when there appears an infinite (percolating)
cluster in a system. Properties  of long flexible polymer chains
in good solvents are described in terms of critical phenomena as
well. One can go on mentioning different examples of phase
transitions and critical phenomena but let me rather attract your
attention to the following question: what is your opinion
regarding the state of the art in this matter and especially
regarding possible future developments? To start, I just want to
mention two polar opinions existing in the field: one of them
states that it is a fascinating part of physics still actively
developed and there is a lot of work to be done there both on the
fundamental level as well as in applications. The other opinion is
that all principal work in the phase transitions theory has
already been done by the middle of the 80-ies and there is neither
new physics there to be explained nor need to apply efforts in
theoretical studies. You are welcome to give your support or
objections to any of these opinions as well as to tell us what in
your mind is happening and will be happening in the field of phase
transitions and critical phenomena.
\begin{figure}[htbp]
\epsfxsize 123mm \epsfysize 70mm
\centerline{\epsffile{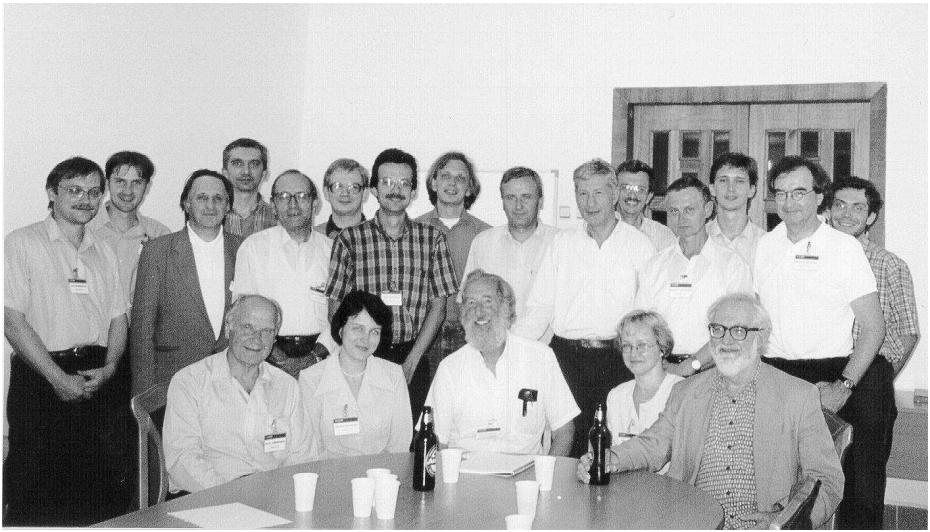}}
\caption{\label{fig1}Participants of the round table ``Phase
transitions and critical phenomena: past, present, and future" (Lviv,
August 30, 2000). Seated (from left to right): I.  Yukhnovskii, O.
Patsahan, M. Schick, A. Ciach, G. Stell. Standing:  Yu. Dublenych, M.
Dudka, R. Levitskii, O. Ivankiv, I. Stasyuk, O.  Velychko, Yu.
Holovatch, C. von Ferber, A. Zagorodnii, O. Bakai, I. Pylyuk, M.
Kozlovskii, R. Sokolovskii, W. Schr\"oer, T.  Yavors'kii.}
\end{figure}

\section*{The future of critical phenomena
is its glorious past!}

\noindent{\bf Michael Schick}. Well, I will start out by playing
devil's advocate. Of the two positions you have outlined, I would
take the second. As you say, the theory of phase transitions was
basically completed by the middle of the eighties. It was a
wonderful time. But the phenomena is now well understood, and I
would not counsel students to go into it. Basically one is now, in
my view, cleaning up details, and that is not the kind of thing
which students should be doing. They should be going into new
areas. Critical phenomena was a fantastically exciting field
TWENTY years ago. In terms of the future, I think it will be like
many other areas of condensed matter physics which have been
studied;  it's been understood, so move on!  I don't think that
one should try to keep students in critical phenomena just because
we had such a good time and learned so much from it. It was
fantastic, it was great! But it is basically well understood. As I
said, if I were to counsel students, I would say that you should
learn this material because it's beautiful. It's a wonderful
subject! But I don't think that's where you should spend your time
in research. You should go into something which is not so well
understood, and which presents new problems, hopefully of the same
order that critical phenomena presented twenty years ago. I think
it would be a real mistake to try and continue in this area.
Obviously I may be wrong. But to summarize, I would say that the
future of critical phenomena is its glorious past.

\section*{You want to find out very basic principles but you find out
how to apply theory to real systems!}

\noindent {\bf Reinhard Folk}. I would like to take the other
point of view. Of course the theory of critical phenomena  is
understood in principal, but I think theory has proceeded to a
stage where you perform quantitative calculations. One wants  to
test theory within high precision. You might say that this is not
so interesting, but on the other hand if you use such a
complicated theory, you have to be sure that it describes the
phenomena precisely. Take as an example the superfluid phase
transition. There you  measure in space the values of the critical
exponents and then compare them with theory. This is similar to
the tests of the theory in elementary particles physics. That is
one point. The other point is, that one uses the RG theory and
equivalent theories for applications in real systems. There are
many problems which are solved only by simple approaches, e.g.
first order approximations and which are not sufficient. As
examples I like to mention crossover effects in complex fluids,
where you have to consider the influence of two different types of
critical behaviour or magnetic systems with impurities. Or one may
study liquids measuring transport coefficients in nonequilibrium
situations with finite heat currents or under shear.  I also think
that there are a lot of problems not solved below $T_c$ because
they are much more complicated than the equivalent problems above
$T_c$. So, in my opinion there is enough space for more work to
do, also it's not of that basic kind that you say. You want to
find out very basic principles but it is also of value to find out
how to apply theory to real systems. So that's another type of
problems in physics but it has its own value.

\section*{The thing is to successfully
marry  the second virial coefficient and scaling
theory.}

\noindent{\bf George Stell}. Well, I think I agree with what
Michael Schick says about critical phenomena, largely, but there was
an implicit assumption in the way you started out, Michael. I mean you
implicitly identify phase transitions and criticality. When I got to
the scene in the early 1960's there were two separate things that were
interesting. First there was the beginning of what became scaling
theory, and homogeneity, and then renormalization-group theory.
Second, there was an independent but equally interesting and
important set of developments in liquid-state theory to predict
with quantitative accuracy the phase transitions associated with
going from gas to a liquid and to a solid. Those are associated with
what happens in regions in the phase diagram that can be very far
away from a critical point.

In other words, the theory of phase
transitions is much broader than the theory of criticality, I
would say. And that gave rise to a problem that proved to be much
more difficult than people initially expected it to be, I think
$-$ the problem of marrying the results of thermodynamic
perturbation theory that had been fine tuned by the early 1970's
to give very good global thermodynamics away from the critical
point to the dictates of critical-point theory, which at about the
same time had been developed to the point at which it could give
very accurate thermodynamics in a very narrow region around the
critical point. The initial engineering-oriented attempts to glue
those two things together were all unsuccessful.

Roughly speaking,
the reason they were unsuccessful goes like this, technically.
Critical theory tells you that you have homogeneity or scaling of
thermodynamics on all length scales. And homogeneity means you
have something that will be true arbitrarily far away from a
critical point, if you take it literally rather than as asymptotic
condition, because homogeneity is intrinsically a global rather
than a local concept. So the difficult thing was shutting off the
homogeneity as one goes away from the critical region. And one HAS
to do this in a successful theory $-$ after all, the second virial
coefficient never heard of homogeneity. The embarrassing thing was
that it turned out to be very difficult to successfully marry the
second virial coefficient and scaling theory, so to speak.

That's something that's been of continuing interest to me and to
several other groups for many years, and only in the last decade or so
it has been successfully coped with by us. For example by Reatto's
Milano group and by my group, in theories that yield both structure
factors as well as thermodynamics, as by other groups who deal only
with the thermodynamics. And it's still a work in progress, because
one wants to deal not only with Argon and the Ising model but with the
criticality of ionic systems and polar fluids and polymers and
colloids and microemulsions etc., etc. Even when one is looking at
what turns out to be an Ising-like critical point in such complex
and soft-matter systems, the special symmetries of these systems
give rise to new experimental twists to what one observes $-$
different sorts of cross-over phenomena and the like. And in some
of them, one has not only liquid-gas Ising-like criticality in
disguise but the possibility of many other different types of
critical or singular points $-$ tricritical points, Lifshitz
points, etc., etc. that one also wants to be able to treat in the
context of globally accurate theories.

Over the last thirty years there has been remarkable progress in
dealing with dense fluids  of complex molecules away from critical
points.  Thirty years ago we knew already how to deal with simple
fluids by introducing hard-sphere reference potentials and then
adding soft tails to them as a perturbation, treating the
hard-sphere fluid by using things like the Percus-Yevick theory
and the Carnahan-Starling approximation.  Now, with about the same
amount of accuracy and almost the same degree of analytic
simplicity, we can treat dense fluids of arbitrarily long
hard-sphere chains, or dendrimers, or star polymers, as long as
they are composed of hard-sphere monomers. And we know how to do a
perturbation theory for adding attractive tails to the
monomer-monomer hard-sphere interaction. We now also have
wonderfully simple and remarkably accurate theories for the
thermodynamics of dense fluids of associating molecules, based on
work by Wertheim and others that enable us to consider the
association of hard-sphere particles into hard molecules,
including the hard-chain polymers mentioned above. In fact, the
development of such theories of association is a specialty of a
number of the people who are here in Lviv.

One thing that has not been developed to a high level of accuracy
is the thermodynamics and structure for reference potentials that
are repulsive inverse-power potentials proportional to relatively
low inverse powers (4 through 9, say) of the distance between
particle centers. Such potentials are appropriate monomer-monomer
reference-system potentials for a number of soft-matter systems,
and the excess thermodynamics of a soft-sphere system depends upon
temperature and density only through a single variable. There was
some nice work done on such models in the 1970's and early 80's
but its use seems to have languished over the last decade compared
to the use of hard-core reference systems.

Another thing that remains rather poorly developed is analytic
theory in two dimensions.  For example, there is no theory of hard
disks as beautifully simple as the Percus-Yevick theory of hard
spheres.  This is because of deep technical reasons: the solutions
of Weiner-Hopf type equations (or, for that matter, of the
Helmholtz equation) have a more singular analytic structure
(involving logarithms) in even dimensions than in odd dimensions.
But I think more can be done on this problem than has been done,
and it would be worth the effort.

Similar statements can be made about binary mixtures of hard
spheres with nonadditive diameters in three dimensions. There have
been some heroic assaults on this problem from time to time, but
more are needed.

\section*{The next century will be the century, when the interest of
scientists will be shifted towards living systems.}

\noindent{\bf Alexander Chalyi}. I would like to use a few minutes
to speak about future applications of the theory of phase
transitions to living systems, namely to the synaptic transmission
or cell-to-cell communication. This phenomenon plays the same role
in living systems as the intermolecular interactions in non-living
systems.  Actually the first paper "Phase Transitions in
Finite-Size Systems and Synaptic  Transmission" on this topic was
reported by me and Dr. L.M.Chernenko at the Les Houches Workshop
"Dynamic Phenomena at Interfaces, Surfaces and Membranes".

The convenient theory of cell-to-cell communication is commonly based
on the ideas concerning chemical intermediaries (transmitter
agents) securing transmission between two neurons in the synaptic
cleft or between the motor neuron and muscle fiber in the
neuro-muscular junction. The sequence of major events in cholinergic
synapse are as follows:  Acetylcholine ($Ach$) - let us consider this
typical transmitter - is synthesized and stored in spheroid vesicles,
then releases and reacts with specific acetylcholine receptors ($R^*$)
in active states. The formation of the transmitter-receptor complex
($Ach - R^*$) produces conformal changes in the postsynaptic  membrane
and therefore the change in a membrane potential. Thus, it is the main
process of transmitting the information from cell to cell, say, in a
human brain.  It is possible to write a scheme of biochemical
reactions and corresponding kinetic equations describing  the temporal
evolution of concentrations of receptors $R^*$ and
transmitter-receptor complexes $Ach-R^*$. It must be stressed that the
process of $Ach$ release is cooperative: about 107 $Ach$ molecules are
released simultaneously under the influence of one nerve impulse.
Furthermore, such synchronous activation of  such a large zone of
receptors by $Ach$ can be considered in detail as the process which
is isomorphic to the critical phenomena in binary liquid mixtures near
the critical mixing point.

So, it's a very important application of the phase transitions
theory. We mostly speak about technical applications like displays on
liquid crystals, etc. Without any doubt, these technical applications
are very useful. Nevertheless, it seems to me that the next
century will be the time when the interests of scientists (first of
all - physicists) will be shifted towards living systems.

And the last remark: the Chairman of the Les Houches Workshop, I
mentioned  above, was a famous French physicist P.G. de Gennes. It was
exactly in February, 1991 and he received his Noble Prize several
months later. In the foreword to the proceedings of this workshop
({\em "Dynamic Phenomena at Interfaces, Surfaces and Membranes". Eds.
D.Beysens, N.Boccara and G.Forgacs, New York, Nova Science Publishers,
1993})  P.G.de Gennes started with the words "Biophysics is a Tower of
Babel" and concluded:  "I left Les Houches with a more optimistic view
of  biophysics. I am in fact convinced that nearly all physicists and
chemists should have an education in biology -- not only in the
molecular aspects, which are usually invoked, but also a panorama of
the living world -- telling us repeatedly that Nature is much more
inventive than us".

\section*{The  future for the phase transitions theory
is the mixed theory: my mind + computer.}

\begin{figure}[htbp]
\epsfxsize 90mm \epsfysize 70mm
 \centerline{\epsffile{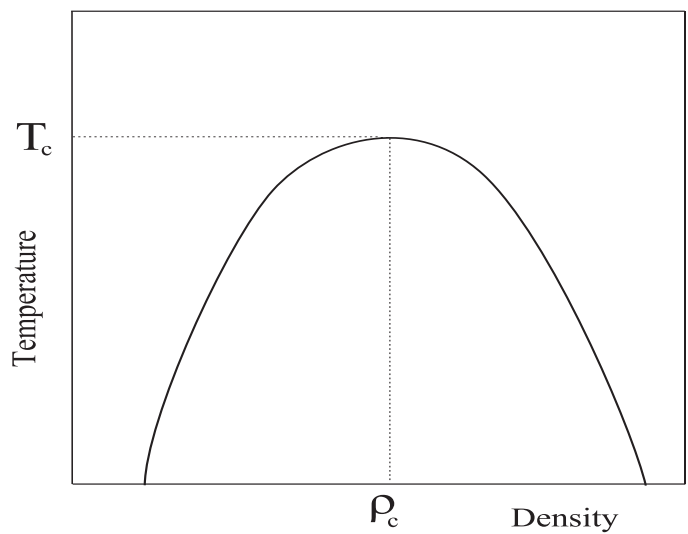}}
\caption{\label{fig2} A coexistance curve and a critical point for
a liquid-gas phase transition.}
\end{figure}

\noindent{\bf Ihor Yukhnovskii}. I would like to tell you about my
view on these different works which were presented at our
conference. You see, when we want to talk about the phase
transition, we, of course, may use the mean field approximation
and obtain this very nice coexistence curve (see Fig. \ref{fig2})
within the  Van der Waals--like theory. We can also obtain the
coexistence curve from numerical simulations but near the critical
point this result might be wrong.
It may be good far from the critical point, but near the
critical point the fluctuations will grow and it will be difficult to reach
it.  The critical exponents  in the critical point
are non-trivial. What's to be done? Is it enough
for us to use the mean field or, knowing that in this point the mean
field is not valid, we have to build a new theory and to consider different
complicated tasks concerning the phase transition on the base of this
theory?  As an example, I would like to very briefly sketch a
very old problem, namely, the liquid-gas critical point. It was the
most difficult task in the phase transition theory.

We have the Hamiltonian with the interaction:
\begin{equation}
U(r_{ij})=\psi(r_{ij})+\phi(r_{ij}),
\end{equation}
consisting  of two parts: $\psi(r_{ij})$ describes repulsive
forces, and $\phi(r_{ij})$ is  the attractive part which,
for example, can be chosen in the form of the Lennard-Jones potential.
Why does the phase transition occur?  The phase transition is
present because of a very long tail of the attraction in the
Lennard-Jones potential.  How shall we describe the phase transition?
Of course, we can construct the free energy containing some additional
variable.  This variable is an  order parameter and we can work with
this free energy.  Having the expression for the free energy we may
talk about the phase transition. But this is a constructive theory,
this theory is not rigorous. The phase transition can be described
rigorously only if we use a proper phase space. We will call it the
collective  variables phase space. Among the variables in this phase
space, there must be a variable which in some limit gives the
macroscopic order parameter. For every task, for every physical
problem there exists its own set of appropriate collective variables
and we have to find such  variables. To give an example, let us
discuss the liquid--gas critical point. You know that in this case the
order parameter is related to density. Therefore, we have to introduce
some variables which are the modes of the density fluctuations:
\begin{equation}
 \hat \rho_{\vec k}=\sum_{i=1}^{N}\exp(-i\vec k\vec r_{i}).
\end{equation}
And the partition function may be written as:
\begin{equation}
\Xi=\Xi_{0}\int(d\rho)\exp[\beta\mu_{1}\rho_{0}-
\frac{\beta}{2V}\sum_{\vec k}\tilde \phi(k)\rho_{\vec k}\rho_{-\vec
k}]J(\rho),
\end{equation}
where $\Xi_{0}$ is the hard sphere contribution: it is the grand partition
function of the hard sphere reference system. The potential
energy of attraction has the square form in the new variables and
$J(\rho)$ is the most difficult part, the Jacobian of the transition from
Cartesian coordinates to these new collective coordinates. Note that among
these coordinates there is one which is connected with the order parameter,
namely, $\hat \rho_{\vec k}$ for $k=0$. We get the following representation for
$J(\rho)$:
\begin{eqnarray} \nonumber
J(\rho) & = &
\int(d\omega)\exp\left
[i2\pi\sum_{\vec k} \omega_{\vec k}\rho_{\vec k}\right]
\exp\left [\sum_{n\geq 1}
\frac{(-i2\pi)^{n}}{n!}\times
\right. \\ \label{jac}
&  & \left.
\sum_{\vec k_{1}\ldots\vec k_{n}}
M_{n}(\vec k_{1},\ldots,\vec k_{n}) \omega_{\vec
k_{1}}\ldots\omega_{\vec k_{n}}\right],
\end{eqnarray}
where $M_{n}(\vec k_{1},\ldots,\vec k_{n})$ is the density cumulant
of $n$th order defined for the reference system.
When you put all cumulants with $n\geq2$ equal to zero you obtain the
analogy of a mean field theory. This means that the  critical exponents
will be wrong, they will be the critical exponents of the
mean field theory.  If you want to obtain right critical exponents in
all the region you should take into account ${\cal M}_n$ with
$n=1,2,3,4$.  And you will obtain the right theory, the right
exponents, and you can obtain the correct thermodynamics.

The order parameter is defined by a collective variable $\rho_{\vec k}$
with $k=0$. The following theorem for the cumulants
$M_{n}(\vec k_{1},\ldots,\vec k_{n})$ (\ref{jac}) can be proved:
when one of the arguments $k$ is equal zero, the
cumulant is automatically equal to zero if the finite size system is
considered. Only when the thermodynamic limit $N,V \rightarrow \infty$
is taken the cumulants ${\cal M}_n$ for $k=0$ are non-zero and can be
expressed by some thermodynamic derivatives. This means that you
should be careful when you study a finite system.

And finally, I want to say that new posibilities in studies of phase
transitions are expected from some "mixed" theory:  my mind + computer, not
only a computer without mind, not only mathematics without computer
because the things are very complicated. And now about different
problems, bioorganic problems and so on. You have seen during our
conference that very complicated systems can be described
by non-central forces. In the collective variables method we can consider
any potential: symmetric, non-symmetric and so on. The theory will be more
complicated but you have a computer, so you can work with such systems.
To conclude, I believe that the most important thing in the theory of phase
transitions is to have the method which is quite general and which allows one
to obtain the correct values for critical exponents and makes
it possible to obtain entire thermodynamic functions.

\noindent{\bf Wolffram Schr\"oer}.
I don't really want to continue the discussion about future of the
critical phenomena. The situation we are facing in respect to the
critical phenomena is to some extent reminiscent to the situation in
quantum mechanics in the late twenties of last century when the basic
equations were developed, however the tedious work to required
understanding of chemistry which is still progressing was not done.
I would rather ask just after you talk, Professor Yukhnovskii. How
much is really done in your method? Have you really got the
non-classical exponents in this method?

\noindent{\bf Ihor Yukhnovskii}. Yes, of course, quite naturally,
without any adjusting parameters, only knowing the  interaction
potential between particles you obtain all thermodynamical
functions, critical exponents. They are of quantitative accuracy.

\noindent{\bf Michael Schick}. The critical exponents in most
phase transitions are known by now with high enough accuracy.

\noindent{\bf Ihor Yukhnovskii}. Yes, we obtain the same critical
exponents, of course.

\noindent{\bf Reinhard Folk}. May I make some comments. The
problem just mentioned by Professor Yukhnovskii is neither to
calculate the value of the asymptotic critical exponents, nor the
Van der Waals theory but it is the crossover between these two
limiting behaviours. You have to find a method which is systematic
and which can give  both regions correct. And there are theories
e.g. by Anisimov and Sengers who calculate this crossover in a
phenomenological way. They can adjust the equation of state in
such a way that they have correct asymptotic exponents together
with  correct amplitude ratios and that crosses over in the
background  to the  classical equation of state. This could be the
Van der Waals equation or any other equation. I think, it would be
very important to prove, that with the method described by
Professor Yukhnovskii you really can do the same. Then you have a
theory which describes the crossover in a systematic way. This
would be an improvement over all theories existing so far.

\noindent{\bf Alexander Chalyi}. I would like to say that I know
the very important results obtained in this method of collective
variables. Rigorous transition from the microscopic Hamiltonian,
or microscopic variables to order parameter variables, to
collective variables through the Jacobian of this transition. This
point is absent in any other approach. Because the starting point
of the renormalization group theory is to use the effective
Hamiltonian. And here the coefficients are obtained explicitly.
This is the first.

The second. Concerning this extended equation of state, it seems
to me, there are different wrong opinions that it must be the
crossover between the so-called scaling and classical behaviour inside
the critical region for reduced temperature variable $(T-T_c)/T_c$
between zero and one. But the existence of such a transition depends
on the value of Ginzburg number. If the Ginzburg number is of order of
one and more there is no region where the crossover occurs. If the
Ginzburg number is smaller than one, then you must have the
crossover.

\noindent{\bf  Yurij Holovatch}. I want to comment that in fact
both of you are right. Because experimentally unobservable
crossover region simply means that it is very narrow and of course
its width depends on the microscopic features of a real system. The
cossover is always present, but its width is system-dependent.

\noindent{\bf Reinhard Folk}. There is already  experimental
evidence, comparing the critical behavior of   $Xe$ and $He^3$,
that one finds within the so-called critical region  deviations
from the renormalization group results. These are caused by the
effects which are background effects. So other lengths  come into
play already in this region and a variety of crossover phenomena
can be observed.

\noindent{\bf Yurij Holovatch}. So, the crossover scenario is
system dependent. Please, Doctor Kozlovskii.

\noindent{\bf  Mykhailo Kozlovskii}. I would like to propose to
your attention some remarks concerning the behaviour of non-universal
quantities at the second order phase transition. In the collective
variables method it is possible to calculate not only universal
properties but non-universal ones. As you know the partition
function can be represented in the following form:
\begin{eqnarray} \label{kozl1}
Z=C e^{a_0N} \int \exp \left( -\frac{1}{2} \sum_{{\bf k}\in {\cal
B}} d(k)\rho_{\bf k}\rho_{ -{\bf k}}-\right.\nonumber\\
\left.-\sum_{n\geq 2} \frac{a_{2n}}{(2n)!}(N)^{1-n} \sum_{{{\bf
k}_1... {\bf k}_{2n}} \atop{{\bf k}_i \in {\cal B}}}\rho_{{\bf
k}_1}...\rho_{{\bf k}_{2n}} \delta_{{\bf k}_1+...+{\bf k}_{2n}}
\right) (d\rho)^{N},
\end{eqnarray}
where $d(k)=a_2-\beta\Phi(k)$, $\Phi(k)$ is the Fourier transform
of the short-range interaction potential, ${\cal B}$ is Brillouin
zone boundary and coefficients $a_n$ are some constants.

The partition function or the free energy has two different types
of contributions: the energy contribution and the contribution
which is  connected with the interaction potential (the entropy
contribution). The first one $\frac{1}{2} \sum_{{\bf k}\leq {\cal
B}} \beta\Phi(k)\rho_{\bf k}\rho_{ -{\bf k}}$ is  diagonal in the
$\rho_{{\bf k}}$ -representation, and the entropy contribution
$\sum_{{\bf l}}\left(\sum_{n\geq 0} \frac{a_{2n}}{(2n)!}
(\rho_{\bf l})^{2n}\right)$ is diagonal in the $\rho_{{\bf
l}}$-representation. So we can not calculate the partition
function exactly. There are two ways, two possible approaches to
calculate Eq. (\ref{kozl1}). The first one uses $\rho_{\bf k}$
variables and is connected with the application of the Gaussian
moments, but in this case we have some condition
$a_{2n}<<(a_2-\beta\Phi(k))^n$, $n=2,3..$ which is right far from
the phase transition point and is not right near the phase
transitions point, because the difference $a_2-\beta\Phi(k)$ turns
to zero when $k\rightarrow 0$ and $T\rightarrow T_c$.

\begin{figure}[htbp]
\epsfxsize 100mm \epsfysize 60mm
 \centerline{\epsffile{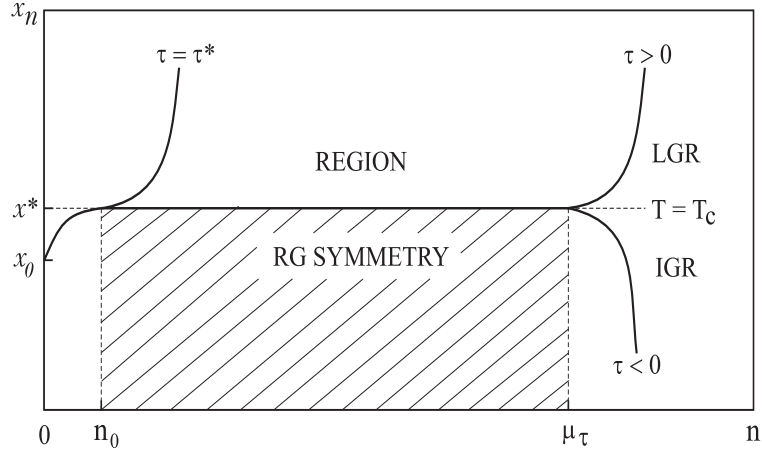}}
\caption{\label{fig3} Evolution of the quantity $x_n
 =  \sqrt 3 <d_n(k)>_{B_{n+1},B_n}/\sqrt{a_n}$ with $n$
for temperatures close to $T_c$.
 }
\end{figure}

\begin{figure}[htbp]
\epsfxsize 100mm \epsfysize 75mm
 \centerline{\epsffile{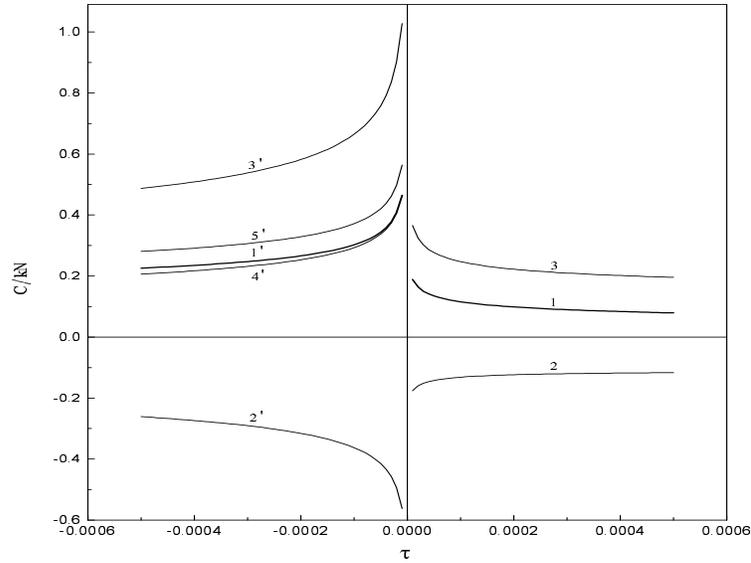}}
\caption{\label{fig4} Heat capacity for small $\tau=(T-T_c)/T_c$.}
\end{figure}

The second way is connected with the
 $\rho_{{\bf l}}$-representation. In this case the entropy contribution
is diagonal and for the energy contributions we use some
approximations (Yukhnovskii approach: see {\em I.R. Yukhnovskii. Phase
Transitions of the Second Order: Collective Variables Method.
World Scientific, Singapore, 1980}). After some calculations we
have obtained the following representation for the partition
function:
\begin{eqnarray} \label{kozl2}
Z&=&CZ_1...Z_{n-1}\int_{R_n} (d\rho)^{N_n} \exp [E_n(\rho)],
\\ \nonumber
E_n&=& -\frac{1}{2} \sum_{{\bf k}\in {\cal B}_n}d_n(k)\rho_{\bf
k}\rho_{ -{\bf k}}- \frac{1}{4!}\frac{1}{N_n} \sum_{{{\bf
k}_1...{\bf k}_{4}}\atop{{\bf k}_i\in {\cal B}_n}}a_n \rho_{{\bf
k}_1}...\rho_{{\bf k}_{4}} \delta_{{\bf k}_1+...+{\bf k}_{4}}.
\end{eqnarray}
The quantity $Z_n$ is the partial partition function for the $N_n$
block structure and $E_n(\rho)$ is the microscopic analogue of
Ginzburg-Landau Hamiltonian. But in the collective variables
method we know the analytic expressions for the coefficients $d_n$
and $a_n$. If we want to calculate, for example, the  free energy
near the phase transition point we must take into account two
different types of contributions. One part is connected with the
renormalization group symmetry region and the other part
corresponds to the limit Gaussian region (see Fig.\ref{fig3}).
Some results of the calculation of the specific heat of 3D one
component model are shown in Fig.\ref{fig4}. You can see that the
value (curve 2) which corresponds to the renormalization group
contribution to the specific heat is negative. The critical
exponent is correct. We have no problems with the calculation of
the critical exponent, but the critical amplitude is negative.
The contribution which corresponds to the limit Gaussian region
(LGR) is shown by the curve 1. The specific heat critical exponent
for this
contribution is the same, but the sign of the critical amplitude
is positive. So, we obtain the resulting curve (3) for the
specific heat which has right critical exponent and right critical
amplitude.

If you want to calculate only universal quantities we can use only
the region where the renormalization group symmetry takes place.
In the case one wants to calculate some non-universal properties,
for instance, critical temperature, the size of the critical
region, the critical amplitude and so on, one has to take into
account the contribution from the limit Gaussian region in the
case $T>T_c$ and inverse Gaussian region  (IGR) for $T<T_c$.

\noindent{\bf Ihor Yukhnovskii}. I would like to attract your
attention. When we calculate the partition function by means of
the quartic distribution function, it has the second degree and
the forth degree of the collective variables of the density. We
have two regions: the Gaussian region and the renormalization
group one (see Fig.\ref{fig3}). In this renormalization group
region we have to use only the quartic distribution function. But
we can not obtain right results when we work only in this region.
The second small region at the end (see Fig.\ref{fig3}) is a limit
Gaussian region. When the temperature turns to the critical
one, this region shrinks to zero. But this region always gives
a very big contribution to the free energy. And the specific heat
becomes positive. This is a very interesting point which I wanted to
show you. So, I would like to repeat that first we must introduce a
very rigorous mathematical apparatus. It can be very complicated, of
course, but we have got a computer. Also we have  a possibility to
obtain the explicit expressions for thermodynamic functions of the
models studied and to investigate their dependence on microscopic
parameters, for example on the interaction parameters.

\noindent{\bf Michael Schick}. What are you interested in? Are you
interested in the criticality {\em per se}? Or you are interested
in non-universal quantities?

\noindent{\bf Ihor Yukhnovskii}. We want to obtain all: an
equation of state, a free energy, an entropy, a susceptibility,
critical indices. And all this is obtained from the first
principles (you have to known only the interaction potential), in
the framework of a unique approach.

Also, I want to tell you what we cannot obtain. We cannot obtain
the solution for the $d=2$ model. Because when $d=2$ I have to
take into account all terms in the Gibbs distribution. This means
that this problem has an exact solution.

\section*{There are possibilities not only to develop the theory
but also to apply it\dots}

\noindent{\bf Alina Ciach}. I don't want to continue this
discussion, but I would like to draw your attention to very
important, I think, and interesting problems which can be solved
by applications of the theory of phase transitions to very complex
systems such as biological systems.

The theory of phase transitions is developed quite well, it's
improving and still developing.  There are, however, interesting
and important problems concerning complex systems, for example
biological systems, which could be solved by applying the theory
of phase transitions in its present form. The structure in
biological cell or organella presumably depends on which phase
should be stable in such complex systems in given conditions, and
the structure determines the function or the phenomena which occur
in such systems. There are even simple questions concerning phase
transitions in complex systems, which are not yet solved. I don't
want to describe the methods which should be used to describe the
phase transitions in real biological systems. However, the
structures formed in biological systems are very similar to the
structures which spontaneously occur in self-assembling systems,
such as lipids in water or surfactants in water or surfactants in
water and oil mixtures, or co-polymers. In such systems solute
particles are amphiphilic, and therefore self-assemble into
 monolayers or bilayers, to prevent unfavourable contacts between polar
and nonpolar particles or particle parts.  These monolayers or
bilayers can arrange in very complex structures in different
conditions. The size of the domains which are surrounded by these
bilayers or monolayers is one or two orders of magnitude larger
than molecular sizes. So, there is another length scale in the problem.
\begin{figure}[htbp]
\epsfxsize 90mm \epsfysize 70mm
 \centerline{\epsffile{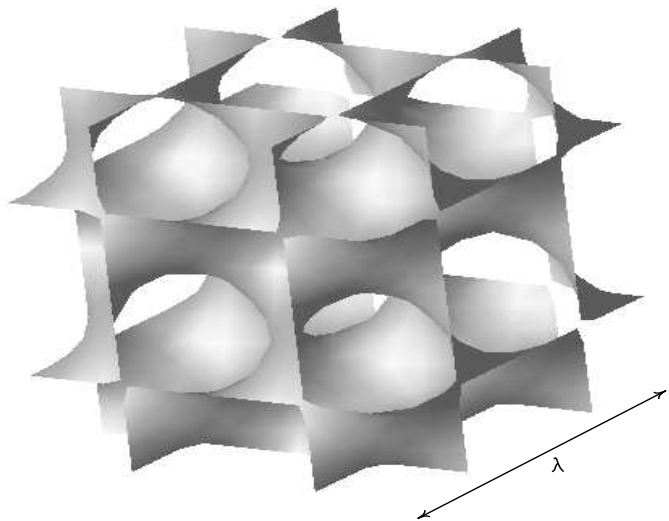}}
\caption{\label{fig5} A unit cell of a cubic diamond phase.}
\end{figure}
An example of such a complex structure is the cubic
diamond phase, which is periodic in space and whose unit cell is
shown in Fig. \ref{fig5}. There are also different periodic
phases, for example the giroid phase shown by Prof. Schick during
his lecture. Endoplasmic reticulum, consisting of a phospholipid
bilayer and extending over a large part of the cell cytoplasm, has
such a complex structure.

The phase behaviour in such systems is studied intensively in the
recent years, however, the effect of surfaces or of confinement on
such systems has not been studied so far very extensively.
Confinement and boundaries certainly play an important role in
biological systems, in particular for organella, which are not
much larger than the distance between the bilayers in the
endoplasmic reticulum.

\begin{figure}[htbp]
\epsfxsize 100mm \epsfysize 60mm
 \centerline{\epsffile{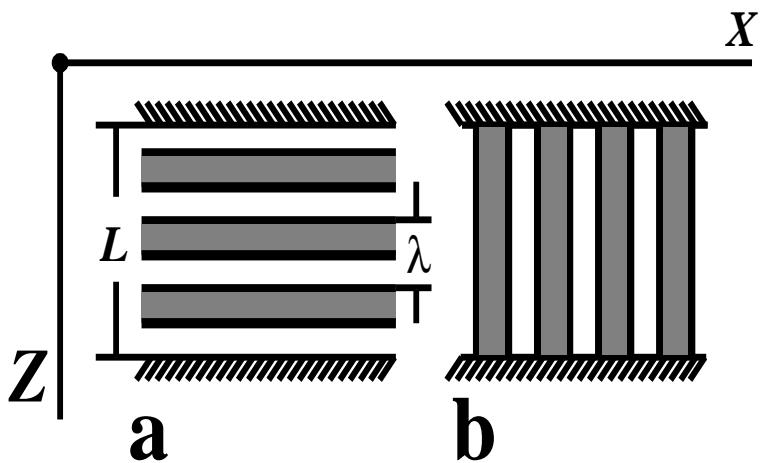}}
\caption{\label{fig6} A lamellar phase confined in a slit with
hydrophilic and neutral walls.}
\end{figure}
Now, what are the effects of various kinds of external surfaces or
of confinement on such self-assembling systems? If we consider a
slit geometry, for example, one can increase the slit size by a
small fraction of the size of the unit cell. In Fig. \ref{fig6} a
lamellar phase confined in a slit with hydrophilic and neutral
walls is shown. When the width $L$ of the slit is not commensurate
with the size of the unit cell $\lambda$, the structure of the
confined ordered phase will be deformed. And the question is, how
this system responds to the stress which is present when $\lambda$
and $L$ do not match? What kinds of deformations will occur?

Condensation of a liquid-like phase  in pores with adsorbing walls
before the bulk phase transition takes place is a well known
phenomenon in simple fluids. Are there similar  capillary
condensation phenomena for ordered periodic phases? How do they
differ from the well known phenomena in uniform systems? The
domains rich in one component and surrounded by amphiphiles  play
on the nanometer length scale a similar role as particles on a
molecular length scale. Therefore some similarities between the
behavior of the simple fluids on the molecular length scale and
self-assembling systems on the nanometer-length scale can be
expected. On the other hand, there should also be differences,
because the domains are not rigid but rather soft and
compressible. Also, symmetry and  anisotropy of  such ordered
phases may affect the phase transitions in confinement. The size
and shape of pores can also have a strong effect on stability of
various phases.  It should be mentioned that unlike in simple
fluids, there are many metastable phases of various structures on
the nanometer length scale in such systems.

Wetting phenomena in such complex systems also deserve attention.
In particular, you may ask whether the wetting phenomena on the
nanometer length scale are the same as in simple fluids on the
molecular length scale, or are  they different? In particular, when
you approach the phase coexistence between the lamellar and cubic
phases, or between the lamellar  and microemulsion phases, what
will be the effect of the chemistry of the wall or of the symmetry
and anisotropy of such phases? There are many questions and I
think they offer large possibilities of applications of the theory
of phase transitions. Even the simplest version of the theory, the
mean field approximation, may give some insight into the problem.
Just by applying the theory of the phase transitions to such
complicated systems, you can describe different phenomena which
have not been described. This is one example of the possibility of
applications of the theory of phase transitions to fascinating
problems, which are close to biological problems. So I liked to
comment that there is not only  a possibility to develop the
theory, but also there is a vast possibility of  its future
applications.

\noindent{\bf Michael Schick}. Also a theory of polymerizing
liquids has huge biological implications. This is  part of
the theory of phase transitions. And we have not touched on
it in any details.

\noindent{\bf Ihor Yukhnovskii}. Biological systems are limited
systems, and there should hold the mean field theory.

\noindent{\bf Michael Schick}. That is certainly true for polymers
(except in the vicinity of a critical point).

\noindent{\bf George Stell}. Critical-point theory can be done by
using the original Wilson approach or by using any one of several
other approaches, as can the crossover problems mentioned by Doctors
Folk, Chalyi, and Holovatch. Among the ones that can be incorporated
into theories that remain useful far from the critical point, there is
a systematic theory by Reatto and Parola and their co-workers.
There is also work by John White, that follows more closely the
original Wilson approach, and there is the development of Anisimov
and Sengers and co-workers, which is more phenomenological but can
be systematized and rigorized. Then there is the SCOZA theory that
I am doing. And there are still others. All of these are different
ways of approaching the same problem that the Lviv group is
interested in. It is easy for each group to get so immersed in its
own formalism that it loses sight of what the other groups are
doing, and to begin to believe that it has THE right way to do things,
rather than A right way. So I think each group has to try to pay
attention to what the others are doing.  Because each of these
approaches has a somewhat different flavour than the others, with its
own set of advantages and disadvantages. So I recommend that from time
to time each group look around and see what the other people are
doing.

\noindent{\bf Alexander Bakai}. Biological systems in operating
state are very sensitive to small changes in temperature,
pressure, osmotic pressure etc. On the other hand they have to
operate stably. And they are open systems which possess a set of
stable dynamic and/or thermodynamic  states. A dynamic system can
change its attractor under action of a small perturbation only if it
is close to an unstable point (e.g. saddle point).  Thermodynamic
system changes its phase state easy in the vicinity of critical point.
In both cases fluctuations of order parameters are considerably large
along with big response functions. How biological systems can combine
a big sensitivity to perturbation (big response functions) with a high
stability (small fluctuations of the order parameters)?

\noindent{\bf Wolffram Schr\"oer}. Biological systems work in
non-equilibrium.

\noindent{\bf Alina Ciach}. I would like to comment on this effect
of large response of a system to a small change of a control
parameter. There are examples of such a behaviour which are not
related to phase transitions, but rather to the structure of these
complex systems.

Consider a lamellar phase like this (Fig. \ref{fig6}a) in a slit
whose walls are strongly hydrophilic. Its orientation is parallel
to the walls. On the other hand,  when the wall is neutral, the
surface tension between the wall and the lamellar phase in
different orientations is almost the same.  However, when the wall
is weakly hydrophilic, the parallel orientation will be preferred
for $\lambda$ and $L$ commensurate, and the perpendicular
orientation will stabilize for slightly different $L$, when the
width of the slit does not fit the period of the lamellar phase.
In both cases the lamellar phase is stable, but the properties  of
the confined system are significantly different.

\noindent{\bf Yurij Holovatch}. Thank you. Following the
discussion I also tried to notice the fields which are
traditionally related to phase transitions and critical phenomena
but which were not mentioned here. Let me name some of them: we
have not touched upon phase transitions on the surface which are
intensively studied now both experimentally and theoretically
(groups of Kurt Binder, Hans Werner Diehl and others). Polymers
were mentioned here. Professor Schick in particular stated that
mean field theory is good in explaining their properties, but this
is when one studies the thermal behaviour and not too close to the
critical point. Speaking about structural properties, in
particular about scaling laws governing these properties one
should rely on  more refined theories such as the renormalization
group approach which is traditionally used there since the
pioneering works of Pierre Gilles de Gennes (let me mention here
schools of Karl Freed, Lothar Sch\"afer). Professor Stell
mentioned a very interesting class of objects: star polymers.
These as well as polymer networks, copolymers and copolymer stars
are the field of intensive studies now. One more generalization of
polymer chains are membranes which themselves form an actively
evolving part of physics. We were not speaking about quantum phase
transitions which occur at zero temperature. A ruling parameter
there is a change in a coupling. Professor Ciach as well as
Professors Chalyi and Bakai mentioned a large response of a system
to a small change of control parameter. This effect was exactly a
reason to introduce one more class of models showing critical
behaviour: the so-called SOC (self-organized criticality) models.
These include forest fire models, traffic jam models etc. As I
have told already the list is far from being complete. Returning
to the start of our discussion: is the phase transition theory
already completed or is it still developing? I think I will
express common opinion if we allow both points of view to exist in
replying to this question.

\section*{Participants of the discussion}

\noindent{\bf Alexander Bakai}, Prof. D.Sc., Kharkiv Institute of
Physics \& Technology, Kharkiv, Ukraine

\noindent{\bf Alina Ciach}, Prof. Dr. hab., Institute of Physical
Chemistry, Polish Acad. Sci., Warszawa, Poland

\noindent{\bf Alexander Chalyi}, Prof. D.Sc., Department of
Physics, O. Bohomolets' National Medical University, Kyiv, Ukraine

\noindent{\bf Reinhard Folk}, a.o.Univ.Prof. Dr. hab., Johannes
Kepler Universit\"at Linz, Linz, Austria

\noindent{\bf Yurij Holovatch}, D.Sc., Institute for Condensed
Matter Physics, National Acad. Sci. of Ukraine, Lviv, Ukraine

\noindent{\bf Mykhajlo Kozlovskii}, D.Sc., Institute for Condensed
Matter Physics, National Acad. Sci. of Ukraine, Lviv, Ukraine

\noindent{\bf Michael Schick}, Prof. Ph.D., Department of Physics,
University of Washington Seattle, Washington, USA

\noindent{\bf Wolffram Schr\"oer}, Prof. Dr. hab., Institut f\"ur
anorganische und physikalische Chemie, Universit\"at Bremen,
Bremen, Germany

\noindent{\bf George Stell}, Prof. Ph.D, Department of Chemistry,
State University of New York at Stony Brook, Stony Brook, USA

\noindent{\bf Ihor Yukhnovskii}, Prof. D.Sc., Institute for
Condensed Matter Physics, National Acad. Sci. of Ukraine, Lviv,
Ukraine

\vspace{4ex} \noindent{\bf Materials prepared by:} Maxym Dudka,
Yurij Holovatch, Roman Melnyk, Ihor Mryglod, Oksana Patsahan
(ICMP), and Taras Yavors'kii (Ivan Franko National University of
Lviv). \label{last@page}
\end{document}